\documentclass[reprint,superscriptaddress,amsmath,amssymb,aps,rmp]{revtex4-2}
\usepackage{graphicx}
\usepackage[dvipsnames]{xcolor}
\usepackage[colorlinks=true,allcolors=NavyBlue]{hyperref}
\setcitestyle{numbers,square,sort&compress}

\makeatletter
\let\cat@comma@active\@empty
\makeatother

\begin{document}

\title{Parallel locomotor control strategies in mice and flies}

\author{Ana I. Gon{\c c}alves}
\thanks{These authors contributed equally.}
\affiliation{Neuroscience Program, Champalimaud Center for the Unknown, Lisbon, Portugal}

\author{Jacob A. Zavatone-Veth}
\thanks{These authors contributed equally.}
\affiliation{Department of Physics, Harvard University, Cambridge, MA, United States}
\affiliation{Center for Brain Science,  Harvard University, Cambridge, MA, United States}

\author{Megan R. Carey}
\thanks{These authors contributed equally; correspondence should be addressed to \href{mailto:megan.carey@neuro.fchampalimaud.org}{megan.carey@neuro.fchampalimaud.org} (MRC) or \href{mailto: damon.clark@yale.edu}{damon.clark@yale.edu} (DAC).}
\affiliation{Neuroscience Program, Champalimaud Center for the Unknown, Lisbon, Portugal}

\author{Damon A. Clark}
\thanks{These authors contributed equally; correspondence should be addressed to \href{mailto:megan.carey@neuro.fchampalimaud.org}{megan.carey@neuro.fchampalimaud.org} (MRC) or \href{mailto: damon.clark@yale.edu}{damon.clark@yale.edu} (DAC).}
\affiliation{Department of Molecular, Cellular and Developmental Biology, Yale University, New Haven, CT, United States}
\affiliation{Department of Physics, Yale University, New Haven, CT, United States}
\affiliation{Department of Neuroscience, Yale University, New Haven, CT, United States}

\date{\today}

\begin{abstract}
Our understanding of the neural basis of locomotor behavior can be informed by careful quantification of animal movement. Classical descriptions of legged locomotion have defined discrete locomotor gaits, characterized by distinct patterns of limb movement. Recent technical advances have enabled increasingly detailed characterization of limb kinematics across many species, imposing tighter constraints on neural control. Here, we highlight striking similarities between coordination patterns observed in two genetic model organisms: the laboratory mouse and \emph{Drosophila}. Both species exhibit continuously-variable coordination patterns with similar low-dimensional structure, suggesting shared principles for limb coordination and descending neural control. 
\end{abstract}

\maketitle

\section*{Introduction}

Locomotion is a fundamental animal behavior. It can be initiated or modulated in response to internal needs, such as thirst, hunger, or other internal states, or in response to external stimuli. Although superficially simple, locomotion requires a large number of muscles to work in coordination to create seemingly effortless movement. This coordinated control must be flexible as well as precise, so that an animal can respond to changes in its environment. The space of possible movement sequences is in principle very high-dimensional, but quantifying animal movement can constrain possible neural solutions to this complex control problem \cite{dickinson2000animals,marder1996principles,krakauer2017neuroscience,marques2018structure,machado2015quantitative,deangelis2019manifold}.

\begin{figure*}[tb]
    \centering
    \includegraphics[width=\textwidth]{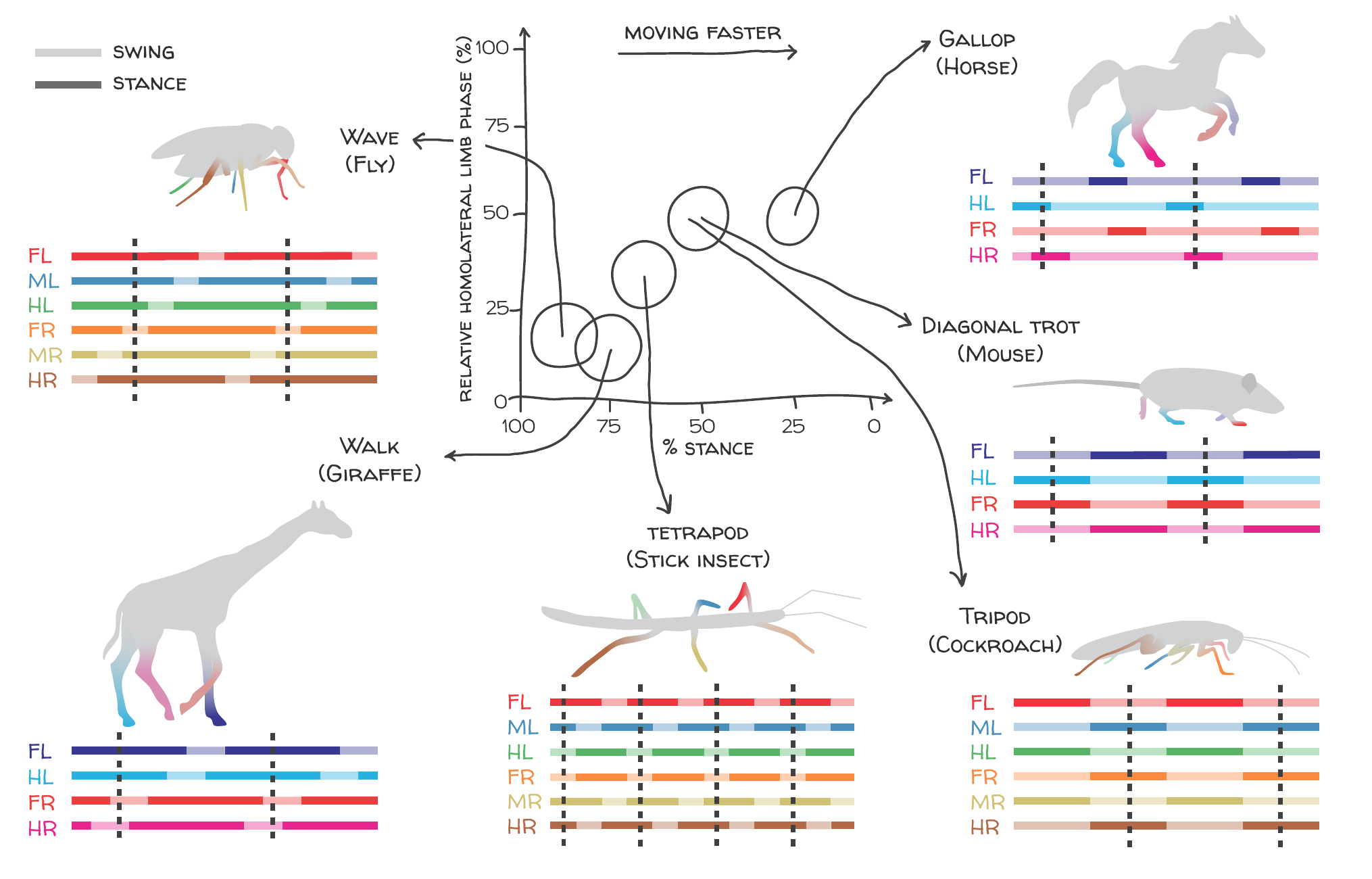}
    \caption{Discrete gaits can be represented by distinct support patterns that depend on the relationship between stance duration and interlimb phasing. In some cases, animals display distinct, preferred patterns of interlimb coordination that can vary depending on size, speed, or species. For example, horses famously alter their gaits at different speeds, with a characteristic gallop at higher speeds. Flies display a wave gait where, at slow speeds, they lift one limb from the ground at a time. Giraffes also have a characteristic slow walk, lifting each limb sequentially. Stick insects display a tetrapod gait, where four limbs touch the ground at each time point in a diagonal arrangement. Cockroaches can show an alternating tripod gait of the six limbs, where diagonal limbs on the ground at the same time. Mice move most of the time in a diagonal trot where one pair of diagonal limbs is in contact with the ground at a time.  This figure is modeled after Figure 5 of \cite{hildebrand1989quadrupedal}, and uses swing-stance patterns from \cite{collins1993coupled,machado2015quantitative,deangelis2019manifold} to estimate the range of relative homolateral limb phases across walking speeds. Canonical stance (solid) and swing (shaded) phases of front, mid- and hindlimbs of the left (FL, ML, HL) and right (FR, MR, HR) sides of the body are illustrated for each species.}
    \label{fig1}
\end{figure*}

Historically, variability in locomotor behavior has often been characterized in terms of discrete motifs \cite{hildebrand1965symmetrical,heglund1974scaling,hoyt1981gait,collins1993coupled,srinivasan2006computer,bellardita2015phenotypic,stephens2008dimensionality,berman2014mapping,marques2018structure}. In legged terrestrial animals, modulation of cyclic stepping patterns with forward speed can occur through transitions between distinct gaits, such as trotting and galloping in quadrupeds \cite{hildebrand1965symmetrical,heglund1974scaling,hoyt1981gait,collins1993coupled,bellardita2015phenotypic,lemieux2016gaits} (Figure \ref{fig1}). Transitions between gaits are characterized by discontinuous changes in limb movement parameters \cite{hoyt1981gait,srinivasan2006computer,nishii2000legged,hildebrand1965symmetrical,hildebrand1989quadrupedal,bidaye2018six,nirody2021universal,collins1993coupled}. The framework of discrete preferred coordination patterns has strongly influenced models for neural control of locomotion \cite{collins1993coupled,aminzare2018gait,aminzare2019heterogeneous,righetti2008pattern,kiehn2016decoding,ausborn2019computational} and analyses of locomotor data \cite{bellardita2015phenotypic,mendes2013quantification,pereira2019fast,bidaye2018six,nirody2021universal} in animals across many phyla. 

Although discrete representations of locomotor patterns can prove useful in summarizing high-dimensional behavior, such representations do not capture the full variability of interlimb coordination patterns. Indeed, even early studies of overground locomotion suggested that patterns may not be truly distinct in all animals, and instead lie along a continuum \cite{hildebrand1965symmetrical,wendler1964laufen}. To accurately constrain neural control mechanisms in different models of animal locomotion, it is therefore critical to fully characterize the variability of their locomotor behaviors.

A complete characterization of locomotor behavior is challenging because the repertoire of these behaviors spans multiple spatial and temporal scales. Modern computer vision techniques can measure the positions of many body parts over time, enabling high-resolution quantification of locomotor kinematics \cite{machado2015quantitative,deangelis2019manifold,mathis2018deeplabcut,nath2019using,pereira2019fast,gosztolai2021liftpose3d,karashchuk2020anipose,dunn2021geometric,hausmann2021measuring}. Here, we highlight striking parallels revealed by these measurement advances in the overground locomotor behaviors of a pair of legged animals: the laboratory mouse \textit{Mus musculus} \cite{machado2015quantitative} and the vinegar fly \textit{Drosophila melanogaster} \cite{deangelis2019manifold}. In both species, inter-limb coordination patterns during spontaneous overground locomotion appear largely continuous, without clear evidence for transitions between discrete patterns. The shared low-dimensional structure of coordination patterns suggests that the neural control requirements in these organisms may be similar \cite{pearson1993common}. 

\section*{Parallel coordination strategies in flies and mice}

\begin{figure}[t]
    \centering
    \includegraphics[width=\columnwidth]{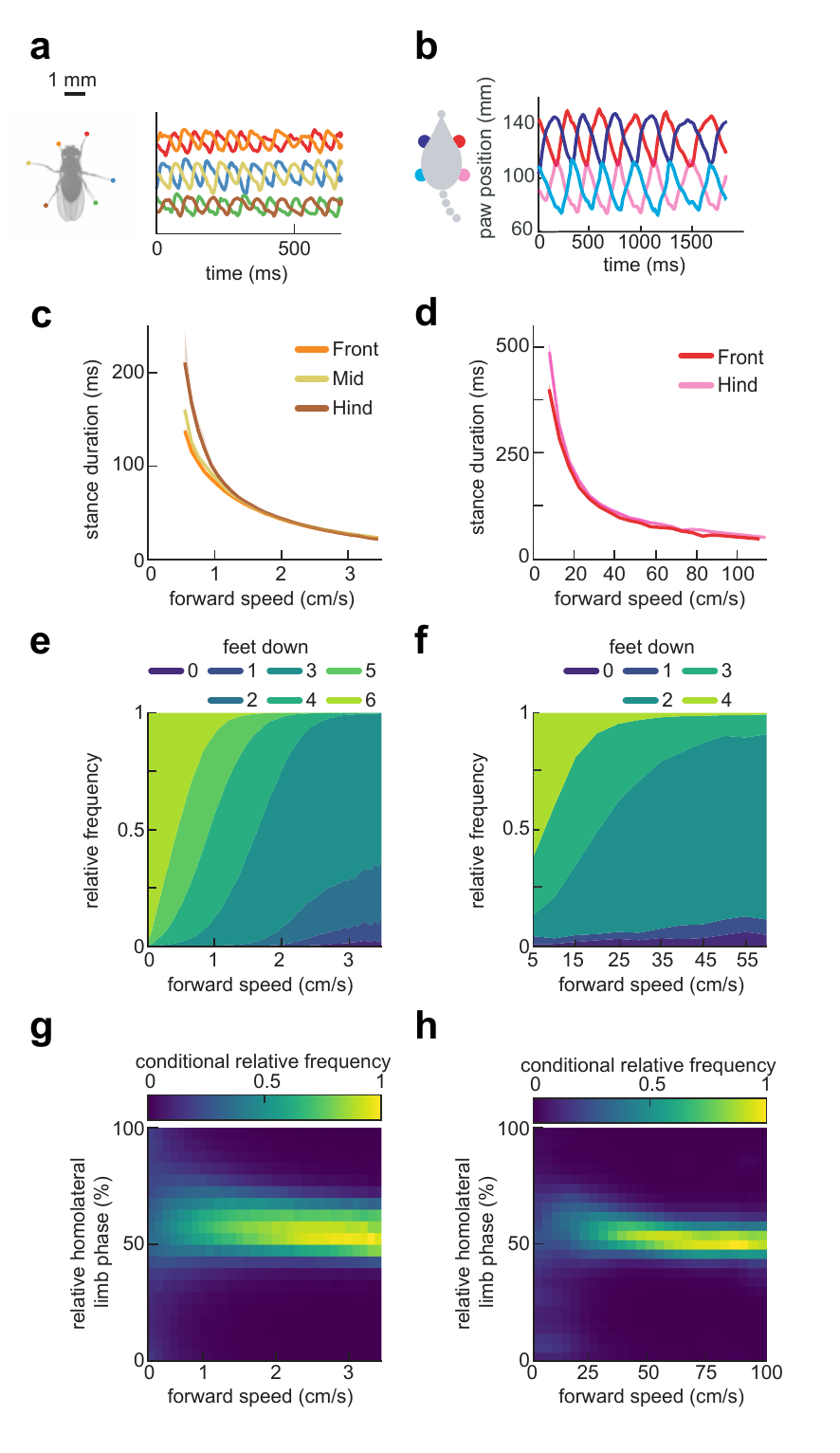}
    \caption{Measurements of fly (left column) and mouse (right column) limb kinematics reveal parallel continua of coordination patterns. (a,b) Continuous forward trajectories over time for a fly's six limb-tips (a) and a mouse's four limb-tips (aka `paws', b). (c,d) Stance duration decreases steeply with forward walking speed. (e,f) Average relative frequencies of limb support patterns within a stride cycle change gradually across forward walking speed for both flies (e) and mice (f; note the expanded speed range). (g,h) Speed-conditioned probability distributions of relative homolateral limb phasing vary smoothly and monotonically with forward walking speed for both flies (g, fore-mid claws) and mice (h, front-hind paws). Fly walking data is adapted from \cite{deangelis2019manifold}; mouse data from \cite{machado2015quantitative}.}
    \label{fig2}
\end{figure}

There are in principle many ways in which an animal could modulate its limb kinematics to regulate speed (Figure \ref{fig1}). Despite this possible degeneracy, the two-dimensional locomotor kinematics of mice and flies spontaneously traversing flat terrain are strikingly similar \cite{machado2015quantitative,deangelis2019manifold}. On average, limb-tip (or paw) kinematic parameters in both flies and mice are smoothly modulated as the animal changes its speed (Figure \ref{fig2}). Stride frequency modulation is achieved mostly by altering the duration of the stance phase of the step cycle, when the limb is in contact with the substrate, rather than the swing phase, when it is lifted and extended (Figure \ref{fig2}c-d) \cite{deangelis2019manifold,machado2015quantitative,strauss1990coordination,mendes2013quantification,wosnitza2013inter,pereira2019fast,chun2021drosophila,nirody2021universal}. In particular, average stance durations are roughly inversely proportional to forward velocity, while average swing durations vary little and stride length increases roughly linearly with speed. These simple analyses suggest that stance duration may be a dominant dimension of kinematic variability \cite{deangelis2019manifold,nirody2021universal,grillner1975locomotion}. 

Basic metrics of inter-limb coordination also vary smoothly with forward speed. A simple way to characterize inter-limb coordination is by the number of supporting limbs that are in stance phase at a given instant, which remains constant for idealized canonical gaits (Figure \ref{fig1}) \cite{collins1993coupled,bellardita2015phenotypic,hildebrand1965symmetrical}. Consistent with the decrease in stance durations with increasing forward speed, the average instantaneous number of limbs in stance phase decreases with increasing speed in both flies and mice (Figure \ref{fig2}e-f). In both animals, this trend reflects speed-dependent enrichment of certain configurations of supporting limbs in different velocity ranges \cite{deangelis2019manifold,machado2015quantitative,strauss1990coordination,mendes2013quantification,wosnitza2013inter,pereira2019fast,chun2021drosophila}. Importantly, the support distributions vary smoothly with speed; there are not sharp transitions between different preferred patterns (Figure \ref{fig2}e-f).

More granular analyses of inter-limb coordination based on relative limb phasing support this common continuum picture. In both flies and mice, distributions of pairwise limb relative phases are unimodal at all forward speeds, with small, smooth monotonic variation in mean relative phases with speed \cite{deangelis2019manifold,machado2015quantitative,szczecinski2018static,chun2021drosophila} (Figure \ref{fig2}g-h). If multiple preferred distinct coordination patterns were used, one would expect these distributions to be multimodal \cite{collins1993coupled}. Thus, although the interlimb coordination pattern of mice, for example, is slightly more walk-like at slower speeds and more trot-like at faster speeds \cite{bellardita2015phenotypic,machado2015quantitative}, there is no categorical boundary, or distinct gait-switching, associated with increasing speed alone. Therefore, observed distributions of relative phasing do not provide evidence that multiple distinct preferred coordination patterns are used by flies or mice during spontaneous locomotion across a wide range of speeds.

\begin{figure*}[t]
    \centering
    \includegraphics[width=\textwidth]{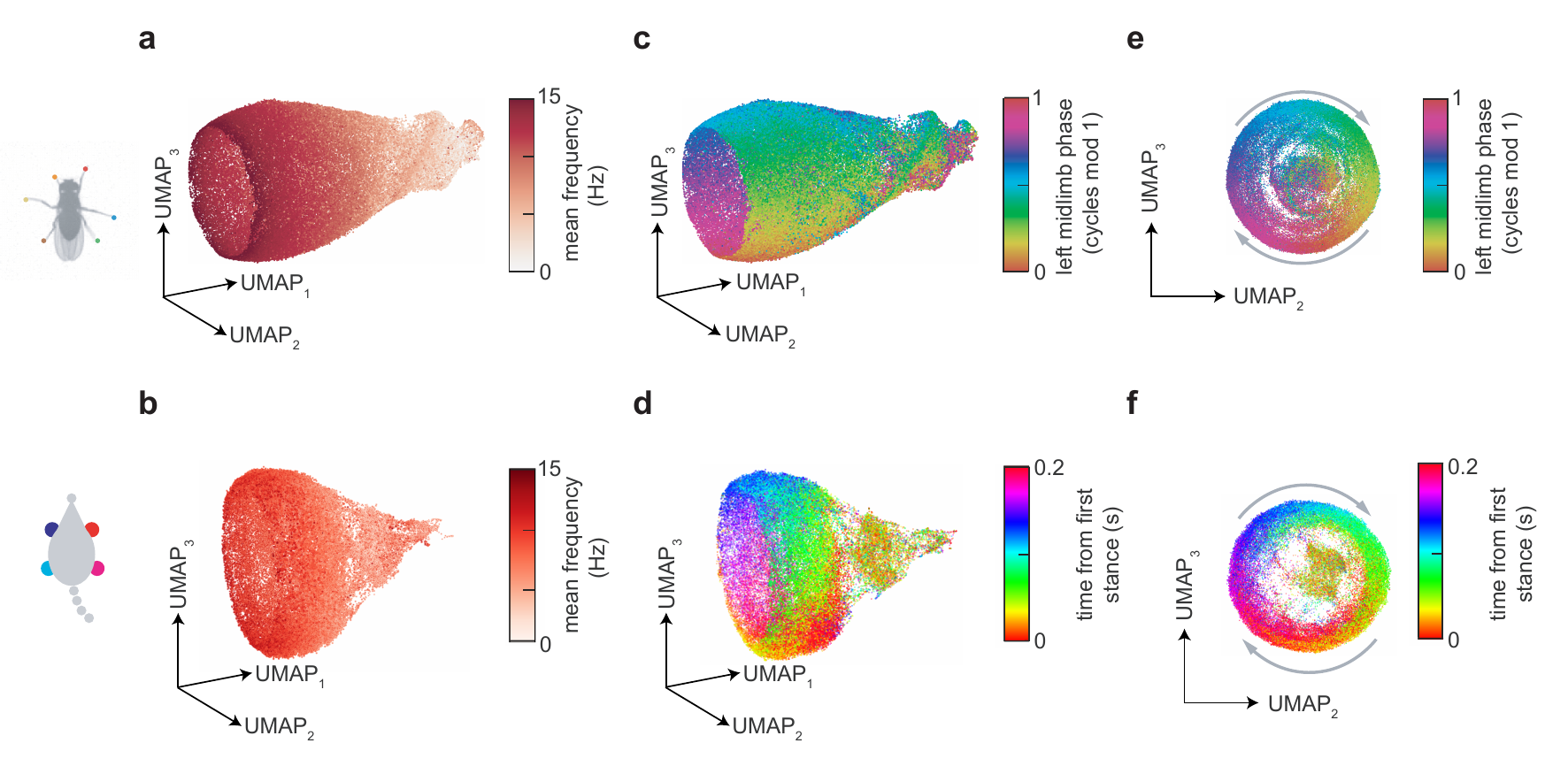}
    \caption{Dimensionality reduction illustrates common low-dimensional structure in fly and mouse interlimb coordination patterns. (a,b) Each point in the UMAP embedding represents a random 200ms segment of limb positions over time, colored by the mean frequency of forward walking for the fly (a) and the mouse (b). (c-f) UMAP embedding of limb kinematic data colored by instantaneous left mid limb phase for the fly (c). For the mouse (d), colors represent the time of the first stance of the front right paw within each segment as a proxy for limb phase to avoid errors in instantaneous phase estimation due to incomplete information about the stride cycle within individual segments. (e-f) same as (c-d) but illustrating the end-on view of the manifold space. Fly limb coordinate time series embeddings were adapted from \cite{deangelis2019manifold}.  Mouse limb positions over time were collected during tied-belt locomotion on a transparent treadmill, as in \cite{darmohray2019spatial}. As in \cite{deangelis2019manifold}, randomly-sampled segments of limb position timeseries were embedded into three dimensions using a Euclidean distance metric and default UMAP hyperparameters. } 
    \label{fig3}
\end{figure*}

Classical metrics of inter-limb coordination show how a subset of features of limb movement change with speed. However, it can be difficult to gain an intuitive understanding for the structure and variability of behavior by manually selecting a small number of features from a high-dimensional dataset. Visualizing such datasets with manifold learning can aid in developing intuition for the structure of behavior \cite{stephens2008dimensionality,berman2014mapping,hausmann2021measuring}. As in DeAngelis \textit{et al.} \cite{deangelis2019manifold}, we used the UMAP algorithm \cite{mcinnes2018umap} to embed segments of fly and mouse limb kinematic timeseries data into three dimensions. In both species, this analysis produced a vase-shaped point cloud, in which the axial dimension corresponds to mean stepping frequency (Figure \ref{fig3}a-b) and the angular dimension corresponds to a global phase (Figure \ref{fig3}c-f). This visualization highlights the similarities between fly and mouse limb coordination strategies, which suggest that they share a common low-dimensional structure. 

Here, we have focused on insights into locomotor coordination resulting from kinematic measurements of limb tips (or paws) of animals traversing flat, featureless terrain. More detailed measurements of legged locomotion in more naturalistic environments and in response to external stimuli will further inform neural control mechanisms. Notably, in larger animals, markerless tracking also enables kinematic measurements in the field \cite{nath2019using}. 

\section*{Implications of parallel coordination strategies for descending neural control}

\begin{figure}
    \centering
    \includegraphics[width=0.8\columnwidth]{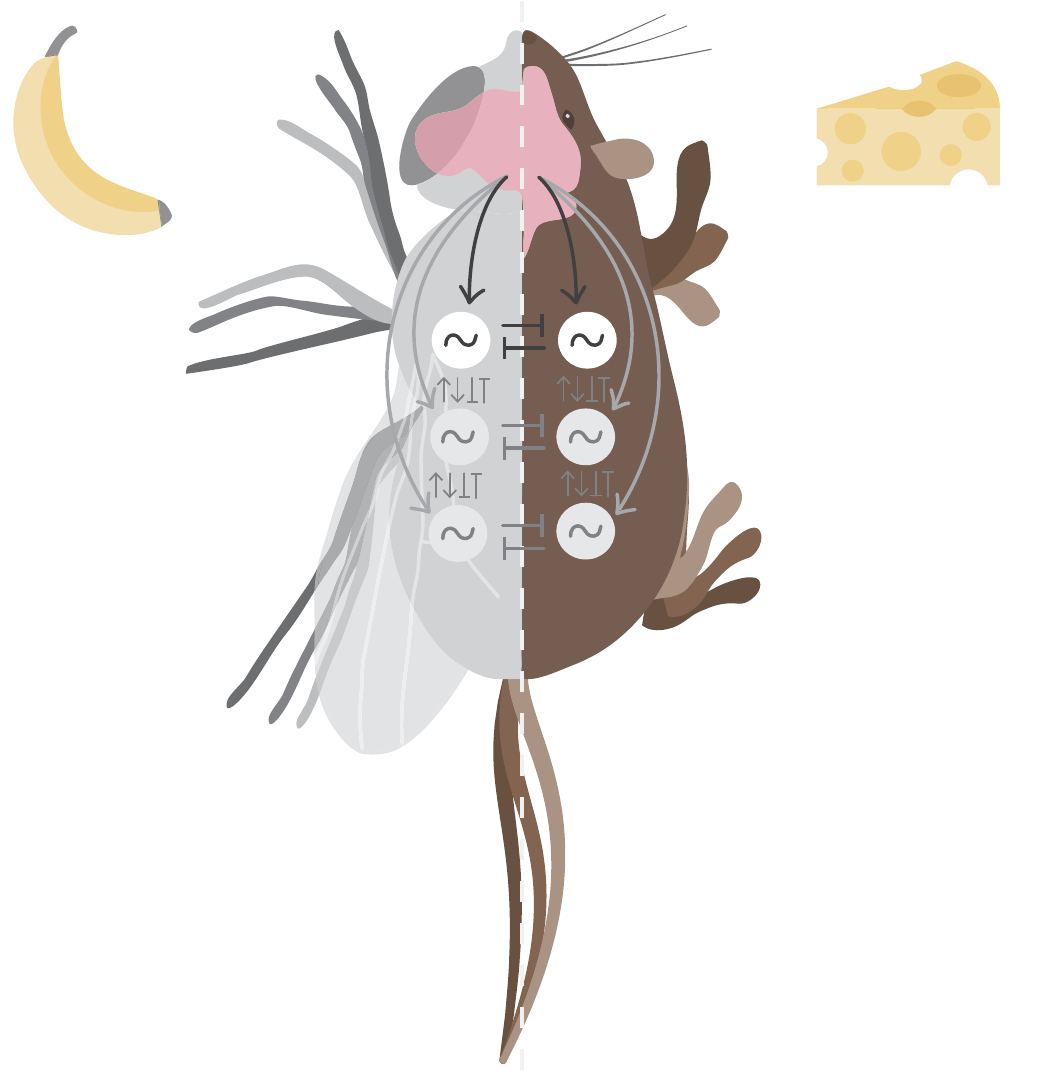}
    \caption{Parallel locomotor control strategies in flies (left) and mice (right). Descending information from the central nervous system, driven by external stimuli, internal state, and/or sensory feedback, modulates locomotor speed and/or direction by modulating CPG modules in the ventral nerve cord of the fly or spinal cord of the mouse, either directly or by modulating internal coupling between CPGs (represented by arrows).}
    \label{fig4}
\end{figure}

The parallel low-dimensional structures of limb coordination in mice and in flies suggest shared principles for neural control of forward speed modulation \cite{pearson1993common}. In both species, the oscillatory patterns of neural activity required to produce rhythmic limb movements are believed to be generated by bilaterally-symmetric central pattern generating circuits (CPGs), in the spinal cord of the mouse or the ventral nerve cord of the fly \cite{bidaye2018six,kiehn2016decoding} (Figure \ref{fig4}). Coordination between limb movements is then established by coupling between CPGs. To smoothly modulate forward speed without causing the animal to deviate from its intended course, these circuits must be modulated symmetrically. The common structure of fly and mouse limb coordination illustrated in Figures \ref{fig2} and \ref{fig3} corresponds exactly to this coupled-oscillator idea: the common frequency of the CPGs is modulated continuously, and their oscillation can be summarized by a single global phase \cite{collins1993coupled,marder1996principles}. Thus, it is possible that one-dimensional command signals could suffice to modulate speed, without the need for detailed descending control of the coupling between CPGs. 

Recent studies have begun to dissect descending neural control of forward speed and movement direction in both flies and mice. In mice, supraspinal areas carry instructions to initiate locomotion \cite{dasilva2018dopamine} in a context-dependent setting \cite{pinto2018connecting}. Similarly, recent work in flies has identified the set of descending neurons that transmit control signals from the central brain to the ventral nerve cord \cite{namiki2018functional}, including individual channels to initiate walking \cite{bidaye2014neuronal,bidaye2020two,cande2018optogenetic}. Neurons involved in the termination of locomotion \cite{bouvier2015descending,capelli2017locomotor,cande2018optogenetic,zacarias2018speed}, speed modulation \cite{caggiano2018midbrain,capelli2017locomotor,bidaye2014neuronal,deangelis2019manifold,cande2018optogenetic}, and steering \cite{cregg2020brainstem,usseglio2020control,bidaye2020two,cande2018optogenetic,rayshubskiy2020neural} have also been identified in both species. Thus, it seems possible that flies and mice may share parallel principles for descending neural control of locomotion. Similar pathways for controlling the speed \cite{severi2014neural} and direction \cite{orger2008control} of locomotion have also been identified in zebrafish, suggesting that these principles may be more broadly conserved. 

In studying how these descending neurons affect changes in limb movements, it will be important to dissect how they modulate the pattern-generating circuitry. In principle, descending commands could directly modulate CPG oscillations \cite{aminzare2018gait,aminzare2019heterogeneous}, directly modulate neural couplings between CPGs \cite{righetti2008pattern}, or indirectly modulate CPGs by altering the gain of sensory feedback. Insects may implement the last of these control strategies \cite{bidaye2018six,berendes2016speed}. In mice there is evidence for all three control strategies. Brain descending inputs send commands to spinal motor circuits modulating locomotor states \cite{pinto2018connecting}, long-distance projection neurons in the spinal cord couple segments regulating forelimbs and hindlimbs, thus modulating interlimb coordination \cite{ruder2016long}, and sensory feedback is also necessary to modulate locomotor cycle and ongoing movement \cite{koch2017rorbeta,fink2014presynaptic,conway1987proprioceptive}. 

Another critical supraspinal structure for coordinated movement in vertebrates is the cerebellum. For locomotion, the cerebellum coordinates movements across the body in space and time, sending continuous calibration signals to the spinal cord to ensure that whole-body coordination is maintained and adapted to changes in the environment  \cite{machado2015quantitative,darmohray2019spatial,machado2020shared,bastian2007cerebellum}. In general, the extensive supraspinal control of mouse and vertebrate locomotor circuits likely bestows additional capabilities for behavioral flexibility in diverse contexts and dynamic environments, beyond the scope of continuous forward walking that we have focused on here.

\section*{Outlook}

Locomotor behavior is well conserved across legged vertebrate species---for instance, despite large differences in body size, mass and limb configurations, mammals and birds exhibit similar kinematic patterns \cite{catavitello2018kinematic}. In this review, we highlighted the shared low-dimensional, continuous patterns of locomotion in freely walking flies and mice, demonstrating that coordination patterns can be similar across phyla. These findings support the idea that similar kinematic principles and neural control mechanisms may underlie walking in these evolutionarily distant species \cite{pearson1993common}.

Recent developments in tracking technology have the potential to provide a more granular description of locomotor kinematics, including 3D tracking of joint angles across the body  \cite{nath2019using,gosztolai2021liftpose3d,karashchuk2020anipose,dunn2021geometric}. More detailed kinematic measurements will provide more stringent constraints on the neural control of locomotion, particularly on the requirements for precise control of individual limbs \cite{karashchuk2020anipose}. 

Beyond locomotor kinematics, it will be exciting to investigate locomotor dynamics, namely, how locomotor forces contribute to the structure and stability of limb coordination patterns. Though some techniques are available \cite{jindrich1999many,jindrich2002dynamic,mendes2014kinematic,dallmann2017load,schiebel2019mechanical,nirody2021tardigrades}, measuring and manipulating dynamics remains comparatively challenging. Moreover, these methods have not yet been applied to high-throughput experiments in model organisms. Developing new techniques to measure the dynamics of legged locomotion in naturalistic environments will be an important step towards revealing the interactions between organism and environment that underlie locomotion. High-resolution EMG recording \cite{zia2020flexible} combined with detailed kinematic analysis is likely to be particularly useful here. 

Paralleling these opportunities for experimental advances, there exist opportunities for new theoretical work on models of central pattern generating circuits. Thus far, many modeling efforts have focused on incorporating CPGs that support the generation of multiple distinct gaits, and analyzing how low-dimensional descending control signals allow the CPG network to switch between those gaits \cite{collins1993coupled,aminzare2018gait,aminzare2019heterogeneous,righetti2008pattern,ausborn2019computational}. As measurements of coordination patterns improve and the underlying neural circuits become better understood, theoretical work will be required to encompass these more complete descriptions of locomotor states. In particular, the observations reviewed here highlight opportunities to develop models that produce continuous sets of inter-limb coordination patterns. 

The continuity of inter-limb coordination patterns observed in \emph{Drosophila} and laboratory mouse (Figures \ref{fig2} and \ref{fig3}) does not imply that all hexapods and quadrupeds share similar control principles. Indeed, many animals exhibit clearly distinct gaits (Figure \ref{fig1}) \cite{collins1993coupled,bidaye2018six,srinivasan2006computer,bellardita2015phenotypic,lemieux2016gaits,marques2018structure}. Moreover, in mice in particular, a broader range of gait patterns can emerge during escape behaviors \cite{bellardita2015phenotypic,caggiano2018midbrain}, or with genetic perturbations \cite{crone2009mice}. As detailed behavioral measurements become possible in a broader range of species, it will be important to carefully characterize the full variability of locomotor behavior in each, without importing assumptions from related animals. Careful dissection of behavior, combined with measurement and manipulation of neural activity and with mathematical modeling, is an essential tool for revealing principles for the neural control of locomotor behavior.

\begin{acknowledgments}
We thank Rita F{\'e}lix for preparing Figures 1 and 4 and Dana Darmohray for comments on an earlier version of the manuscript. This work was supported by a fellowship (PD/BD/128291/2017) from the Portuguese Fundação para a Ciência e a Tecnologia (to AIG) and grants from the Simons-Emory International Consortium on Motor Control (Simons Foundation \#717106), the European Research Council (866237), and the Fundação para a Ciência e a Tecnologia (PTDC/MED-NEU/30890/2017) to MRC. JAZ-V was supported by the NSF-Simons Center for Mathematical and Statistical Analysis of Biology at Harvard and the Harvard Quantitative Biology Initiative (award \#1764269), as well as the Harvard FAS Dean's Competitive Fund for Promising Scholarship. DAC was supported in this work by the Smith Foundation and NIH R01 EY026555. The authors declare no conflict of interest.
\end{acknowledgments}

\section*{References}

\noindent 
Papers of particular interest, published within the period of review, have been highlighted as:

\textbf{*} of special interest

\textbf{**} of outstanding interest 

\bibliographystyle{elsarticle-num}
\bibliography{refs}

\begin{thebibliography}{10}
\expandafter\ifx\csname url\endcsname\relax
  \def\url#1{\texttt{#1}}\fi
\expandafter\ifx\csname urlprefix\endcsname\relax\def\urlprefix{URL }\fi
\expandafter\ifx\csname href\endcsname\relax
  \def\href#1#2{#2} \def\path#1{#1}\fi

\bibitem{dickinson2000animals}
M.~H. Dickinson, C.~T. Farley, R.~J. Full, M.~Koehl, R.~Kram, S.~Lehman, How
  animals move: an integrative view, Science 288~(5463) (2000) 100--106.

\bibitem{marder1996principles}
E.~Marder, R.~L. Calabrese, Principles of rhythmic motor pattern generation,
  Physiological Reviews 76~(3) (1996) 687--717.

\bibitem{krakauer2017neuroscience}
J.~W. Krakauer, A.~A. Ghazanfar, A.~Gomez-Marin, M.~A. MacIver, D.~Poeppel,
  Neuroscience needs behavior: correcting a reductionist bias, Neuron 93~(3)
  (2017) 480--490.

\bibitem{marques2018structure}
J.~C. Marques, S.~Lackner, R.~F{\'e}lix, M.~B. Orger, Structure of the
  zebrafish locomotor repertoire revealed with unsupervised behavioral
  clustering, Current Biology 28~(2) (2018) 181--195.

\bibitem{machado2015quantitative}
A.~S. Machado, D.~M. Darmohray, J.~Fayad, H.~G. Marques, M.~R. Carey, A
  quantitative framework for whole-body coordination reveals specific deficits
  in freely walking ataxic mice, eLife 4 (2015) e07892, \\\textbf{**By
  developing a tracking system to measure limb, nose and tail kinematics, the
  authors showed whole-body coordination deficits in Purkinje cell degeneration
  mice. This approach and results provides a framework for measuring cerebellar
  ataxia deficits and, consequently, the role of cerebellum in predicting the
  consequences of movement.}

\bibitem{deangelis2019manifold}
B.~D. DeAngelis, J.~A. Zavatone-Veth, D.~A. Clark, The manifold structure of
  limb coordination in walking {Drosophila}, eLife 8 (2019) e46409,
  \\\textbf{**This paper characterized the two-dimensional limb-tip kinematics
  of a large number of freely-walking \emph{Drosophila}. They showed that
  variability in coordination patterns appears continuous, without clear
  evidence for distinct preferred patterns. Moreover, they showed that
  modulation of speed in response to neural perturbations or visual stimulation
  appears to occur mostly through modulation of stance duration.}

\bibitem{hildebrand1965symmetrical}
M.~Hildebrand, Symmetrical gaits of horses, Science 150~(3697) (1965) 701--708.

\bibitem{heglund1974scaling}
N.~C. Heglund, C.~R. Taylor, T.~A. McMahon, Scaling stride frequency and gait
  to animal size: mice to horses, Science 186~(4169) (1974) 1112--1113.

\bibitem{hoyt1981gait}
D.~F. Hoyt, C.~R. Taylor, Gait and the energetics of locomotion in horses,
  Nature 292~(5820) (1981) 239--240.

\bibitem{collins1993coupled}
J.~J. Collins, I.~N. Stewart, Coupled nonlinear oscillators and the symmetries
  of animal gaits, Journal of Nonlinear Science 3~(1) (1993) 349--392.

\bibitem{srinivasan2006computer}
M.~Srinivasan, A.~Ruina, Computer optimization of a minimal biped model
  discovers walking and running, Nature 439~(7072) (2006) 72--75.

\bibitem{bellardita2015phenotypic}
C.~Bellardita, O.~Kiehn, Phenotypic characterization of speed-associated gait
  changes in mice reveals modular organization of locomotor networks, Current
  Biology 25~(11) (2015) 1426--1436.

\bibitem{stephens2008dimensionality}
G.~J. Stephens, B.~Johnson-Kerner, W.~Bialek, W.~S. Ryu, Dimensionality and
  dynamics in the behavior of {C.} elegans, PLoS Computational Biology 4~(4)
  (2008) e1000028.

\bibitem{berman2014mapping}
G.~J. Berman, D.~M. Choi, W.~Bialek, J.~W. Shaevitz, Mapping the stereotyped
  behaviour of freely moving fruit flies, Journal of The Royal Society
  Interface 11~(99) (2014) 20140672.

\bibitem{lemieux2016gaits}
M.~Lemieux, N.~Josset, M.~Roussel, S.~Courad, F.~Bretzner, Speed-dependent
  modulation of the locomotor behavior in adult mice reveals attractor and
  transitional gaits, frontiers in Neuroscience 10~(42).

\bibitem{nishii2000legged}
J.~Nishii, Legged insects select the optimal locomotor pattern based on the
  energetic cost, Biological Cybernetics 83~(5) (2000) 435--442.

\bibitem{hildebrand1989quadrupedal}
M.~Hildebrand, The quadrupedal gaits of vertebrates, Bioscience 39~(11) (1989)
  766.

\bibitem{bidaye2018six}
S.~S. Bidaye, T.~Bockem{\"u}hl, A.~B{\"u}schges, Six-legged walking in insects:
  how {CPGs}, peripheral feedback, and descending signals generate coordinated
  and adaptive motor rhythms, Journal of Neurophysiology 119~(2) (2018)
  459--475.

\bibitem{nirody2021universal}
J.~A. Nirody, Universal features in panarthropod inter-limb coordination during
  forward walking., Integrative and Comparative Biology.

\bibitem{aminzare2018gait}
Z.~Aminzare, V.~Srivastava, P.~Holmes, Gait transitions in a phase oscillator
  model of an insect central pattern generator, SIAM Journal on Applied
  Dynamical Systems 17~(1) (2018) 626--671.

\bibitem{aminzare2019heterogeneous}
Z.~Aminzare, P.~Holmes, Heterogeneous inputs to central pattern generators can
  shape insect gaits, SIAM Journal on Applied Dynamical Systems 18~(2) (2019)
  1037--1059.

\bibitem{righetti2008pattern}
L.~Righetti, A.~J. Ijspeert, Pattern generators with sensory feedback for the
  control of quadruped locomotion, in: 2008 IEEE International Conference on
  Robotics and Automation, IEEE, 2008, pp. 819--824.

\bibitem{kiehn2016decoding}
O.~Kiehn, Decoding the organization of spinal circuits that control locomotion,
  Nature Reviews Neuroscience 17~(4) (2016) 224--238.

\bibitem{ausborn2019computational}
J.~Ausborn, N.~A. Shevtsova, V.~Caggiano, S.~M. Danner, I.~A. Rybak,
  Computational modeling of brainstem circuits controlling locomotor frequency
  and gait, eLife 8 (2019) e43587.

\bibitem{mendes2013quantification}
C.~S. Mendes, I.~Bartos, T.~Akay, S.~M{\'a}rka, R.~S. Mann, Quantification of
  gait parameters in freely walking wild type and sensory deprived {Drosophila}
  melanogaster, eLife 2 (2013) e00231.

\bibitem{pereira2019fast}
T.~D. Pereira, D.~E. Aldarondo, L.~Willmore, M.~Kislin, S.~S.-H. Wang,
  M.~Murthy, J.~W. Shaevitz, Fast animal pose estimation using deep neural
  networks, Nature Methods 16~(1) (2019) 117--125, \\\textbf{*This paper
  presents LEAP, an animal pose estimation method based on deep learning
  techniques that is able to describe the position of multiple body parts. It
  uses deep convolution networks and it requires a small labelling dataset.}

\bibitem{wendler1964laufen}
G.~Wendler, Laufen und stehen der stabheuschrecke {Carausius} morosus:
  sinnesborstenfelder in den beingelenken als glieder von regelkreisen,
  Zeitschrift f{\"u}r vergleichende Physiologie 48~(2) (1964) 198--250.

\bibitem{mathis2018deeplabcut}
A.~Mathis, P.~Mamidanna, K.~M. Cury, T.~Abe, V.~N. Murthy, M.~W. Mathis,
  M.~Bethge, {DeepLabCut}: markerless pose estimation of user-defined body
  parts with deep learning, Nature Neuroscience 21~(9) (2018) 1281--1289,
  \\\textbf{*This paper developed a videography-based tracking system that uses
  transfer learning with deep neural networks to estimate body part kinematics.
  It is a very versatile method that works on a variety of behaviors and across
  many species.}

\bibitem{nath2019using}
T.~Nath, A.~Mathis, A.~C. Chen, A.~Patel, M.~Bethge, M.~W. Mathis, Using
  {DeepLabCut} for {3D} markerless pose estimation across species and
  behaviors, Nature Protocols 14~(7) (2019) 2152--2176.

\bibitem{gosztolai2021liftpose3d}
A.~Gosztolai, S.~G{\"u}nel, V.~Lobato-R{\'\i}os, M.~Pietro~Abrate, D.~Morales,
  H.~Rhodin, P.~Fua, P.~Ramdya, {LiftPose3D}, a deep learning-based approach
  for transforming two-dimensional to three-dimensional poses in laboratory
  animals, Nature Methods 18~(8) (2021) 975--981, \\\textbf{*LiftPose3D is a
  method for 3D pose estimation from a single-camera two-dimensional pose
  estimation that uses deep learning techniques. This method does not require
  extensive calibration procedures to estimate three-dimensional pose in
  freely-behaving animals.}

\bibitem{karashchuk2020anipose}
P.~Karashchuk, K.~L. Rupp, E.~S. Dickinson, E.~Sanders, E.~Azim, B.~W. Brunton,
  J.~C. Tuthill, Anipose: a toolkit for robust markerless {3D} pose estimation,
  BioRxiv\\\textbf{*Anipose is a Python-based software to estimate 3D animal
  pose using calibration procedures, filters to resolve tracking errors and a
  triangulation module. The users can transform their 2D tracking results into
  3D pose estimations.}

\bibitem{dunn2021geometric}
T.~W. Dunn, J.~D. Marshall, K.~S. Severson, D.~E. Aldarondo, D.~G. Hildebrand,
  S.~N. Chettih, W.~L. Wang, A.~J. Gellis, D.~E. Carlson, D.~Aronov, W.~A.
  Freiwald, F.~Wang, B.~P. {\"O}lveczky, Geometric deep learning enables {3D}
  kinematic profiling across species and environments, Nature Methods 18~(5)
  (2021) 564--573.

\bibitem{hausmann2021measuring}
S.~B. Hausmann, A.~M. Vargas, A.~Mathis, M.~W. Mathis, Measuring and modeling
  the motor system with machine learning, Current Opinion in Neurobiology 70
  (2021) 11--23.

\bibitem{pearson1993common}
K.~Pearson, Common principles of motor control in vertebrates and
  invertebrates, Annual Review of Neuroscience 16~(1) (1993) 265--297.

\bibitem{strauss1990coordination}
R.~Strauss, M.~Heisenberg, Coordination of legs during straight walking and
  turning in {Drosophila} melanogaster, Journal of Comparative Physiology A
  167~(3) (1990) 403--412.

\bibitem{wosnitza2013inter}
A.~Wosnitza, T.~Bockem{\"u}hl, M.~D{\"u}bbert, H.~Scholz, A.~B{\"u}schges,
  Inter-leg coordination in the control of walking speed in {Drosophila},
  Journal of Experimental Biology 216~(3) (2013) 480--491.

\bibitem{chun2021drosophila}
C.~Chun, T.~Biswas, V.~Bhandawat, Drosophila uses a tripod gait across all
  walking speeds, and the geometry of the tripod is important for speed
  control, eLife 10 (2021) e65878.

\bibitem{grillner1975locomotion}
S.~Grillner, Locomotion in vertebrates: central mechanisms and reflex
  interaction, Physiological Reviews 55~(2) (1975) 247--304.

\bibitem{szczecinski2018static}
N.~S. Szczecinski, T.~Bockem{\"u}hl, A.~S. Chockley, A.~B{\"u}schges, Static
  stability predicts the continuum of interleg coordination patterns in
  {Drosophila}, Journal of Experimental Biology 221~(22) (2018) jeb189142.

\bibitem{darmohray2019spatial}
D.~M. Darmohray, J.~R. Jacobs, H.~G. Marques, M.~R. Carey, Spatial and temporal
  locomotor learning in mouse cerebellum, Neuron 102~(1) (2019) 217--231.

\bibitem{mcinnes2018umap}
L.~McInnes, J.~Healy, J.~Melville, {UMAP}: Uniform manifold approximation and
  projection for dimension reduction, arXiv preprint arXiv:1802.03426.

\bibitem{dasilva2018dopamine}
J.~A. da~Silva, F.~Tecuapetla, V.~Paix{\~a}o, R.~M. Costa, Dopamine neuron
  activity before action initiation gates and invigorates future movements,
  Nature 554~(7691) (2018) 244--248.

\bibitem{pinto2018connecting}
M.~J. Ferreira-Pinto, L.~Ruder, P.~Capelli, S.~Arber, Connecting circuits for
  supraspinal control of locomotion, Neuron 100~(2) (2018) 361--374.

\bibitem{namiki2018functional}
S.~Namiki, M.~H. Dickinson, A.~M. Wong, W.~Korff, G.~M. Card, The functional
  organization of descending sensory-motor pathways in {Drosophila}, eLife 7
  (2018) e34272, \\\textbf{**This research mapped the organization of
  descending neurons that transmit information from the central brain to the
  ventral nerve cord of \emph{Drosophila}. They generated transgenic lines
  targeting specific descending neuron cell types, enabling future functional
  study.}

\bibitem{bidaye2014neuronal}
S.~S. Bidaye, C.~Machacek, Y.~Wu, B.~J. Dickson, Neuronal control of
  {Drosophila} walking direction, Science 344~(6179) (2014) 97--101.

\bibitem{bidaye2020two}
S.~S. Bidaye, M.~Laturney, A.~K. Chang, Y.~Liu, T.~Bockem{\"u}hl,
  A.~B{\"u}schges, K.~Scott, Two brain pathways initiate distinct forward
  walking programs in {Drosophila}, Neuron 108~(3) (2020) 469--485,
  \\\textbf{**This research screened \emph{Drosophila} descending and
  protocerebral neurons to identify pathways that initiate forward walking.
  They identified distinct neuron types that drive initiation of
  object-directed curve walking and rapid straight walking.}

\bibitem{cande2018optogenetic}
J.~Cande, S.~Namiki, J.~Qiu, W.~Korff, G.~M. Card, J.~W. Shaevitz, D.~L. Stern,
  G.~J. Berman, Optogenetic dissection of descending behavioral control in
  {Drosophila}, eLife 7 (2018) e34275.

\bibitem{bouvier2015descending}
J.~Bouvier, V.~Caggiano, R.~Leiras, V.~Caldeira, C.~Bellardita, K.~Balueva,
  A.~Fuchs, O.~Kiehn, Descending command neurons in the brainstem that halt
  locomotion, Cell 163~(5) (2015) 1191--1203.

\bibitem{capelli2017locomotor}
P.~Capelli, C.~Pivetta, M.~S. Esposito, S.~Arber, Locomotor speed control
  circuits in the caudal brainstem, Nature 551~(7680) (2017) 373--377,
  \\\textbf{**This research comprehensively studied the excitatory and
  inhibitory neuronal populations in the caudal brainstem, showing how
  glutamatergic neurons support high-speed locomotion and glycinergic neurons
  lead to a locomotor halt.}

\bibitem{zacarias2018speed}
R.~Zacarias, S.~Namiki, G.~M. Card, M.~L. Vasconcelos, M.~A. Moita, Speed
  dependent descending control of freezing behavior in {Drosophila}
  melanogaster, Nature communications 9~(1) (2018) 1--11.

\bibitem{caggiano2018midbrain}
V.~Caggiano, R.~Leiras, H.~Go{\~n}i-Erro, D.~Masini, C.~Bellardita, J.~Bouvier,
  V.~Caldeira, G.~Fisone, O.~Kiehn, Midbrain circuits that set locomotor speed
  and gait selection, Nature 553~(7689) (2018) 455--460, \\\textbf{**Using
  cell-specific optogenetic targeting, the authors mapped the role of
  excitatory nucleus in the cuneiform nucleus (CnF) and pedunculopontine
  nucleus (PPN), two structures in the midbrain region. Neurons in the CnF
  elicit high-speed synchronous locomotion, whereas neurons in the PPN promote
  slower alternating locomotion.}

\bibitem{cregg2020brainstem}
J.~M. Cregg, R.~Leiras, A.~Montalant, P.~Wanken, I.~R. Wickersham, O.~Kiehn,
  Brainstem neurons that command mammalian locomotor asymmetries, Nature
  Neuroscience 23~(6) (2020) 730--740.

\bibitem{usseglio2020control}
G.~Usseglio, E.~Gatier, A.~Heuz{\'e}, C.~H{\'e}rent, J.~Bouvier, Control of
  orienting movements and locomotion by projection-defined subsets of brainstem
  v2a neurons, Current Biology 30~(23) (2020) 4665--4681.

\bibitem{rayshubskiy2020neural}
A.~Rayshubskiy, S.~L. Holtz, I.~D’Alessandro, A.~A. Li, Q.~X. Vanderbeck,
  I.~S. Haber, P.~W. Gibb, R.~I. Wilson, Neural circuit mechanisms for steering
  control in walking {Drosophila}, bioRxiv.

\bibitem{severi2014neural}
K.~E. Severi, R.~Portugues, J.~C. Marques, D.~M. O’Malley, M.~B. Orger,
  F.~Engert, Neural control and modulation of swimming speed in the larval
  zebrafish, Neuron 83~(3) (2014) 692--707.

\bibitem{orger2008control}
M.~B. Orger, A.~R. Kampff, K.~E. Severi, J.~H. Bollmann, F.~Engert, Control of
  visually guided behavior by distinct populations of spinal projection
  neurons, Nature Neuroscience 11~(3) (2008) 327--333.

\bibitem{berendes2016speed}
V.~Berendes, S.~N. Zill, A.~B{\"u}schges, T.~Bockem{\"u}hl, Speed-dependent
  interplay between local pattern-generating activity and sensory signals
  during walking in {Drosophila}, Journal of Experimental Biology 219~(23)
  (2016) 3781--3793.

\bibitem{ruder2016long}
L.~Ruder, A.~Takeoka, S.~Arber, Long-distance descending spinal neurons ensure
  quadrupedal locomotor stability, Neuron 92~(5) (2016) 1063--1078.

\bibitem{koch2017rorbeta}
S.~C. Koch, M.~G. Del~Barrio, A.~Dalet, G.~Gatto, T.~G{\"u}nther, J.~Zhang,
  B.~Seidler, D.~Saur, R.~Sch{\"u}le, M.~Goulding, {ROR$\beta$} spinal
  interneurons gate sensory transmission during locomotion to secure a fluid
  walking gait, Neuron 96~(6) (2017) 1419--1431.

\bibitem{fink2014presynaptic}
A.~J. Fink, K.~R. Croce, Z.~J. Huang, L.~Abbott, T.~M. Jessell, E.~Azim,
  Presynaptic inhibition of spinal sensory feedback ensures smooth movement,
  Nature 509~(7498) (2014) 43--48.

\bibitem{conway1987proprioceptive}
B.~Conway, H.~Hultborn, O.~Kiehn, Proprioceptive input resets central locomotor
  rhythm in the spinal cat, Experimental Brain Research 68~(3) (1987) 643--656.

\bibitem{machado2020shared}
A.~S. Machado, H.~G. Marques, D.~F. Duarte, D.~M. Darmohray, M.~R. Carey,
  Shared and specific signatures of locomotor ataxia in mutant mice, eLife 9
  (2020) e55356.

\bibitem{bastian2007cerebellum}
S.~Morton, A.~Bastian, Mechanisms of cerebellar gait ataxia, The Cerebellum 16
  (2007) 79--86.

\bibitem{catavitello2018kinematic}
G.~Catavitello, Y.~Ivanenko, F.~Lacquaniti, A kinematic synergy for terrestrial
  locomotion shared by mammals and birds, eLife 7 (2018) e38190.

\bibitem{jindrich1999many}
D.~L. Jindrich, R.~J. Full, Many-legged maneuverability: dynamics of turning in
  hexapods, Journal of Experimental Biology 202~(12) (1999) 1603--1623.

\bibitem{jindrich2002dynamic}
D.~L. Jindrich, R.~J. Full, Dynamic stabilization of rapid hexapedal
  locomotion, Journal of Experimental Biology 205~(18) (2002) 2803--2823.

\bibitem{mendes2014kinematic}
C.~S. Mendes, S.~V. Rajendren, I.~Bartos, S.~M{\'a}rka, R.~S. Mann, Kinematic
  responses to changes in walking orientation and gravitational load in
  {Drosophila} melanogaster, PloS one 9~(10) (2014) e109204.

\bibitem{dallmann2017load}
C.~J. Dallmann, T.~Hoinville, V.~D{\"u}rr, J.~Schmitz, A load-based mechanism
  for inter-leg coordination in insects, Proceedings of the Royal Society B:
  Biological Sciences 284~(1868) (2017) 20171755.

\bibitem{schiebel2019mechanical}
P.~E. Schiebel, J.~M. Rieser, A.~M. Hubbard, L.~Chen, D.~Z. Rocklin, D.~I.
  Goldman, Mechanical diffraction reveals the role of passive dynamics in a
  slithering snake, Proceedings of the National Academy of Sciences 116~(11)
  (2019) 4798--4803.

\bibitem{nirody2021tardigrades}
J.~A. Nirody, L.~A. Duran, D.~Johnston, D.~J. Cohen, Tardigrades exhibit robust
  interlimb coordination across walking speeds and terrains, Proceedings of the
  National Academy of Sciences 118~(35) (2021) e2107289118, \\\textbf{*This
  research showed that tardigrades employ a continuum of inter-limb
  coordination patterns similar to that observed in \emph{Drosophila}. Robust
  coordination is maintained across walking substrates of widely varying
  stiffness.}

\bibitem{zia2020flexible}
M.~Zia, B.~Chung, S.~Sober, M.~S. Bakir, Flexible multielectrode arrays with
  2-d and 3-d contacts for \textit{in-vivo} eectromyography recording, IEEE
  Transactions on Components, Packaging and Manufacturing Technology 10~(2)
  (2020) 197--202.

\bibitem{crone2009mice}
S.~A. Crone, G.~Zhong, R.~Harris-Warrick, K.~Sharma, In mice lacking v2a
  interneurons, gait depends on speed of locomotion, Journal of Neuroscience
  29~(21) (2009) 7098--7109.

\end{thebibliography}

\end{document}